# Modulation instability, Akhmediev Breathers and continuous wave supercontinuum generation

**J. M. Dudley[1]\*, G. Genty[2], F. Dias[3], B. Kibler[4], N. Akhmediev[5]**

[1] *Département d'Optique P. M. Duffieux, Institut FEMTO-ST*
*UMR 6174 CNRS-Université de Franche-Comté, 25030 Besançon, France*

[2] *Optics Laboratory, Department of Physics, Tampere University of Technology, FIN-33101 Tampere, Finland*

[3] *Centre de Mathématique et de Leurs Applications (CMLA), ENS Cachan, France*

[4] *Institut Carnot de Bourgogne, UMR 5209 CNRS/Université de Bourgogne, 21078 Dijon, France*

[5] *Optical Sciences Group, Research School of Physics and Engineering, Institute of Advanced Studies, The Australian National University, Canberra ACT 0200, Australia*

*\*Corresponding author: john.dudley@univ-fcomte.fr*

**Abstract**

Numerical simulations of the onset phase of continuous wave supercontinuum generation from modulation instability show that the structure of the field as it develops can be interpreted in terms of the properties of Akhmediev Breathers. Numerical and analytical results are compared with experimental measurements of spectral broadening in photonic crystal fiber using nanosecond pulses.

# 1. Introduction

Modulation instability (MI) is a characteristic feature of a wide class of nonlinear dispersive systems, associated with the dynamical growth and evolution of periodic perturbations on a continuous wave background [1,2]. In the initial stage of evolution, the spectral sidebands associated with the instability experience exponential amplification at the expense of the pump, but the subsequent wave dynamics is more complex and various scenarios of energy exchange between the spectral modes are possible. This dynamics is closely related to the celebrated Fermi-Pasta-Ulam (FPU) recurrence phenomenon, and several important studies in an optical context have been reported [3-5].

In optics, MI has been extensively studied in the context of fiber propagation as described by the nonlinear Schrödinger equation (NLSE). This has been motivated both by interest in the fundamental dynamics relating to FPU recurrence [3], as well as by potential applications in ultrashort pulse train generation and parametric amplification [6-8]. The spontaneous development of MI seeded from noise has also been shown to underlie the initial stages of fiber supercontinuum generation seeded by continuous wave (CW) radiation or picosecond or nanosecond pulses [9-11]. Links between this phase of spontaneous MI and the development of "optical rogue wave" instabilities have also recently been proposed [12,13].

Somewhat surprisingly, although exact solutions describing MI dynamics have existed in the mathematical physics literature for nearly 25 years [14], subsequent studies have either developed similar ideas independently [15,16] or have focused on approximate truncated or purely numerical approaches [17-21]. However, there is increasing appreciation that the results of Ref. [14] provide a powerful framework with which to interpret a wide range of MI-related dynamics, and the particular class of solution exhibiting FPU-like growth-return evolution is now widely referred to within the fluid dynamics community as the Akhmediev Breather (AB) [22-25]. In this paper, we focus on particular applications in optics, highlighting a number of "rediscovered" features of these analytic breather solutions in the context of current studies of MI dynamics and supercontinuum generation.

Specifically, we use numerical simulations to examine the evolution of a modulated CW field as it nonlinearly compresses to a periodic train of ultrashort pulses, and we confirm that the AB theory quantitatively describes the "maximally-compressed" pulse characteristics over a wide parameter range. We then show that the AB theory also provides significant insight into the properties of the initial phase of CW supercontinuum generation seeded by noise-driven MI. We discuss how the temporal structure that develops from spontaneous MI can be naturally interpreted in terms of the characteristics of AB evolution, and we show in particular how the frequency-domain properties of the maximally-compressed AB solution reproduce the characteristic triangular shape of the MI spectrum when plotted on a semi-logarithmic scale. Numerical and analytic results are confirmed by experimental studies of quasi-CW spectral broadening induced by nanosecond pulse pumping in the anomalous dispersion regime of a photonic crystal fiber (PCF). Our experiments are carried out at fixed fiber length, but the quantitative agreement between the slope of the spectral wings of the MI spectrum and the analytic AB spectrum nonetheless provides evidence for the presence of breather dynamics. These results thus complement more direct studies of dynamical evolution measuring energy exchange between sidebands as a function of power or length [3,26]. Finally, we briefly discuss directions for future research that may potentially link these results with rogue wave instabilities.

**2. Theoretical summary and the breather dynamics of modulation instability**

We first review the theory of Ref. 14 and present a series of simulation results that illustrate its application to describing induced MI. We write the NLSE in dimensional form [27]:

$$i\frac{\partial A}{\partial z} + \frac{\beta_2}{2}\frac{\partial^2 A}{\partial T^2} + \gamma |A|^2 A = 0, \qquad (1)$$

where $|A|^2$ has dimensions of instantaneous power in W and the dispersion and nonlinearity coefficients $\beta_2$ ($< 0$) and $\gamma$ have dimensions of $ps^2$ $km^{-1}$ and $W^{-1}$ $km^{-1}$ respectively. The Akhmediev Breather is an exact analytic solution describing the evolution with $z$ of a wave with initial constant amplitude on which is superimposed a small periodic perturbation taking the form of a *T*-dependent modulation with amplitude that is a fraction of that of the CW field. The solution consists of an evolving train of ultrashort pulses that is periodic in time (*T*) and that exhibits a FPU-like growth-return cycle in propagation distance (*z*). In contrast to soliton solutions which are localized in *T*, the ideal FPU growth-return behavior localizes the breather solution in the *z* direction. As we shall see below, certain initial conditions can also yield periodic evolution along the *z*-direction, but even in this case, the pulse train characteristics nonetheless remain well-described by the analytic AB solution.

The AB solution to Eq. (1) is written explicitly in the form [14]:

$$A(z,T) = \sqrt{P_0}\ \frac{(1-4a)\cosh(bz) + ib\sinh(bz) + \sqrt{2a}\cos(\omega_{mod}T)}{\sqrt{2a}\cos(\omega_{mod}T) - \cosh(bz)}, \qquad (2)$$

which shows growth-return evolution over $-\infty < z < \infty$. Equation (2) represents a family of solutions with a variable independent parameter $\omega_{mod}$ which is a perturbation frequency (of the initial temporal modulation). The coefficients *a* and *b* depend on $\omega_{mod}$ and are defined by: $2a = [1 - (\omega_{mod}/\omega_c)^2]$ and $b = [8a(1-2a)]^{1/2}$ with $\omega_c^2 = 4\gamma P_0/|\beta_2|$ and $P_0$ the power of the CW field at large $|z|$. The solution is valid over the range of modulation frequencies that experience MI gain: $\omega_c > \omega_{mod} > 0$. The coefficient *a* then varies in the interval $0 < a < 1/2$ while the parameter $b > 0$ governs the MI growth. The maximum gain condition $b = 1$ occurs for $a = 1/4$, i.e. $\omega_{mod} = \omega_c/\sqrt{2}$. Note that these conditions that follow from the properties of Eq. (2) are identical to the well-known results obtained using linear stability analysis [2, 27]. The solution in Eq. (2) describes an evolving periodic train of ultrashort pulses with temporal period $T_{mod} = 2\pi/\omega_{mod}$. The individual temporal peaks or sub-pulses have maximum amplitude and minimum temporal width at z = 0. The solution at this point describes the "maximally-compressed" AB, and is given by:

$$A(z=0,T) = \sqrt{P_0}\ \frac{(1-4a) + \sqrt{2a}\cos(\omega_{mod}T)}{\sqrt{2a}\cos(\omega_{mod}T) - 1}. \qquad (3)$$

The nonlinear evolution of a modulated CW wave can be studied in general by solving the NLSE numerically for an input field of the form $A(z=z_0,T) = \sqrt{P_0}\ [1 + \alpha_{mod} \cos(\omega_{mod} T)]$, where the modulation parameter $\alpha_{mod}$ is in general a small complex number. When $\alpha_{mod}$ depends on $\omega_{mod}$ and takes the particular form $\alpha_{mod} = \mu\ [1+ i(2b/\omega_{mod}^2)]$ with $\mu$ real, the evolution is described by the exact result in Eq. (2) which sweeps a heteroclinic orbit in terms of nonlinear dynamics. Any other form for the parameter $\alpha_{mod}$ does not lead to an ideal return to steady state and decay to a plane wave, but rather yields a more complex solution that lies near the heteroclinic orbit with periodicity along the $z$-direction; that is, we see multiple growth-return cycles and periodic energy exchange between sidebands. Albeit more complex, this evolution can still be fully described analytically in terms of Jacobi elliptic functions (see Eq. (18) of Ref. [14]; See also [28-31]).

On the other hand, extensive simulations of the non-exact case with real-valued $\alpha_{mod}$ as initial condition show that Eqs (2) and (3) nonetheless accurately describe the evolution over at least the first growth-return cycle and, moreover, the pulse train at the point of maximum initial temporal compression can be predicted near-exactly by the analytic solution in Eq. (2) for modulation amplitudes over a large range $0 < \alpha_{mod} < 0.25$. This is an important observation, because real-valued initial modulation amplitudes correspond to conditions that are more straightforward to achieve experimentally using dual-frequency excitation [32]. Figs 1 and 2 illustrate this explicitly by showing a selection of simulation results where the NLSE is integrated numerically using the initial conditions described above and for real-valued $\alpha_{mod}$. The simulations use standard single mode fiber parameters at 1550 nm with $\beta_2 = -20$ ps$^2$ km$^{-1}$, $\gamma = 1.1$ W$^{-1}$ km$^{-1}$, and we assume $P_0 = 30$ W. With these parameters, MI gain is observed for modulation frequencies $f_{mod} = \omega_{mod}/2\pi$ in the range $0 - f_c$ where $f_c = \omega_c/2\pi = 408.88$ GHz. Maximum gain at $a = 0.25$ corresponds to modulation frequency $f_{mod} = f_c / \sqrt{2} = 289.12$ GHz. In contrast to Eqs (2) and (3) above, these figures are plotted with the origin $z = 0$ corresponding to the point of injection of the initial field, and the propagation coordinate $z$ is scaled relative to the characteristic nonlinear length $L_{NL} = (\gamma P_0)^{-1} = 30.3$ m.

Firstly, in Fig. 1, we consider a fixed modulation frequency $f_{mod} = 289.12$ GHz at the peak of the MI gain, but we vary the modulation parameter $\alpha_{mod}$. We use a grayscale representation to show the evolution of both the temporal intensity $|A(z,T)|^2$ and spectral intensity $|\tilde{A}(z,f)|^2$, where $A(z,T) \leftrightarrow \tilde{A}(z,f)$ are Fourier transform pairs. The evolution is shown over at least one growth-return cycle along $z$ and the graphs below the grayscale figures compare the shape of the pulse train at the point of maximal temporal compression and spectral expansion (indicated by arrows) with the predicted maximally-compressed AB solution given by Eq. (3). The horizontal axes on the grayscale plots are normalized relative to $f_{mod}$. The results show how increasing $\alpha_{mod}$ decreases the characteristic distance over which periodic growth-return occurs. Figure 1(c) also illustrates how disagreement between numerical simulations and the predictions of the AB theory begins to become apparent as the modulation amplitude increases beyond $\alpha_{mod} = 0.25$.

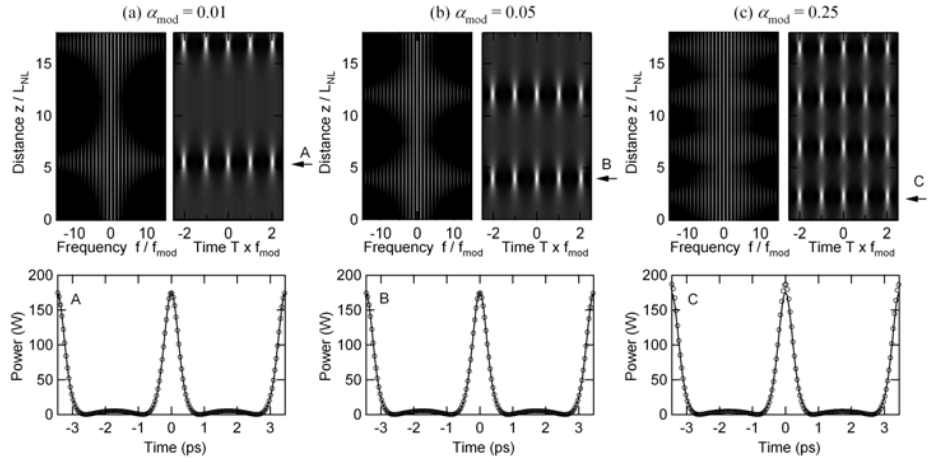

**Fig. 1** Grayscale plots show simulated spectral and temporal evolution for $a = 0.25$; $f_{mod} = 289.12$ GHz and with modulation amplitudes $\alpha_{mod}$ as shown. Arrows and labels indicate the point of maximum temporal compression during the first growth-return cycle. Graphs A, B, C below compare simulation results (circles) with analytic results describing the maximally-compressed AB sub-pulses (lines) from Eq. (3).

Figure 2 shows a second set of results. Here we choose a fixed modulation amplitude of $\alpha_{mod} = 0.01$ (where the AB theory provides an accurate description of the dynamics), but we vary the parameter $a$ to span a range of modulation frequencies $f_{mod}$. We can thus see how the qualitative form of the AB solution changes across the MI gain band. Specifically, Fig. 2 (a) shows the minimal reshaping observed as the frequency approaches the upper limit of MI gain, whilst Figs 2(b) and (c) show cases of decreasing modulation frequency, which would approach the Peregrine soliton in the limit $a = 0.5$ [31]. The evolution is plotted so as to highlight the growth up to the first point of maximal temporal compression. In interpreting Fig. 2, note that the characteristic distance of the initial growth phase up to this point itself depends on $f_{mod}$, asymptotically increasing to $\infty$ at $a = 0$ and $a = 0.5$, whilst attaining its minimal value at the peak of the MI gain when $a = 0.25$; recall that the evolution for this latter case is shown in Fig 1(a).

The results in Figs 1 and 2 show impressive agreement between the properties of the temporal sub-pulses at the point of initial compression extracted from simulations, and the analytic predictions describing the maximally-compressed AB calculated from Eq. (3). This agreement has been confirmed over a range of modulation amplitudes and for various modulation frequencies across the MI gain curve. At first sight, however, this may seem surprising, however, because the case of a purely real-valued initial modulation amplitude does not precisely correspond to the initial conditions of the analytic breather solution derived in Ref. [14]. But in this context, it is important to note that whilst one can obtain an exact solution of the NLSE for each arbitrary initial condition with purely real or imaginary modulation amplitude, all the periodic solutions nonetheless lie very close to a heteroclinic orbit when we move along the trajectories far from the saddle point of the nonlinear phase space [14]. There is thus significant robustness in the ability of the analytic AB solution to describe the initial compression phase for a very wide range of initial conditions.

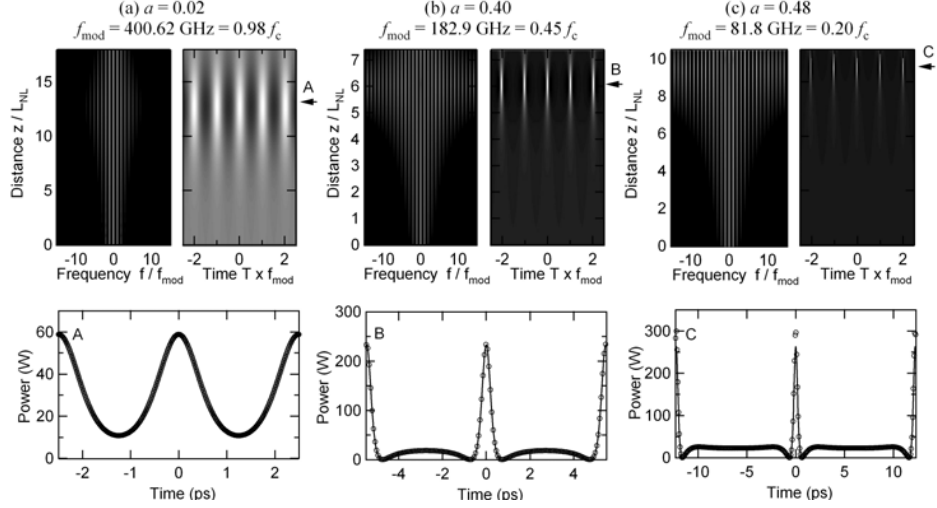

**Fig. 2** Grayscale plots show simulated spectral and temporal evolution for fixed $\alpha_{mod} = 0.01$ but varying modulation parameters ($a$ and thus $f_{mod}$) as shown. Arrows and labels indicate the point of maximum temporal compression during the first growth-return cycle. Graphs A, B, C below compare simulation results (circles) with analytic results describing the maximally-compressed AB sub-pulses (lines) from Eq. (3).

## 3. Akhmediev Breather dynamics and the onset of supercontinuum generation

It is well-known that MI plays a central role in the initial dynamics of quasi-CW supercontinuum generation in the anomalous dispersion regime [10]. In this section we now consider how this onset dynamics can be interpreted in terms of the properties of Akhmediev Breathers. We begin by showing experimental results: Fig. 3(a) shows spectral measurements at the output of 3.9 m of a highly nonlinear PCF for various peak powers using 1 ns pulses at 1064 nm. The fused silica fiber used was the same as that employed in the experiments of supercontinuum generation by Ranka et al. [33], and its structure and parameters have been extensively studied (see for example Ref. [10]).

The experimental results are compared with fully realistic generalised NLSE (GNLSE) simulations shown in Fig. 3(b), assuming 1 ns Gaussian input pulses. Note that the source linewidth (> 10 GHz) of the laser used in these experiments (a passively Q-switched microchip laser) was such that stimulated Brillouin scattering could be neglected in the simulations. A broadband one photon per mode noise background, Raman gain, higher-order dispersion and higher-order nonlinear terms are all included [10]. The simulations average over 10 realisations and are convolved with a spectral resolution function to match experiments (see caption.)

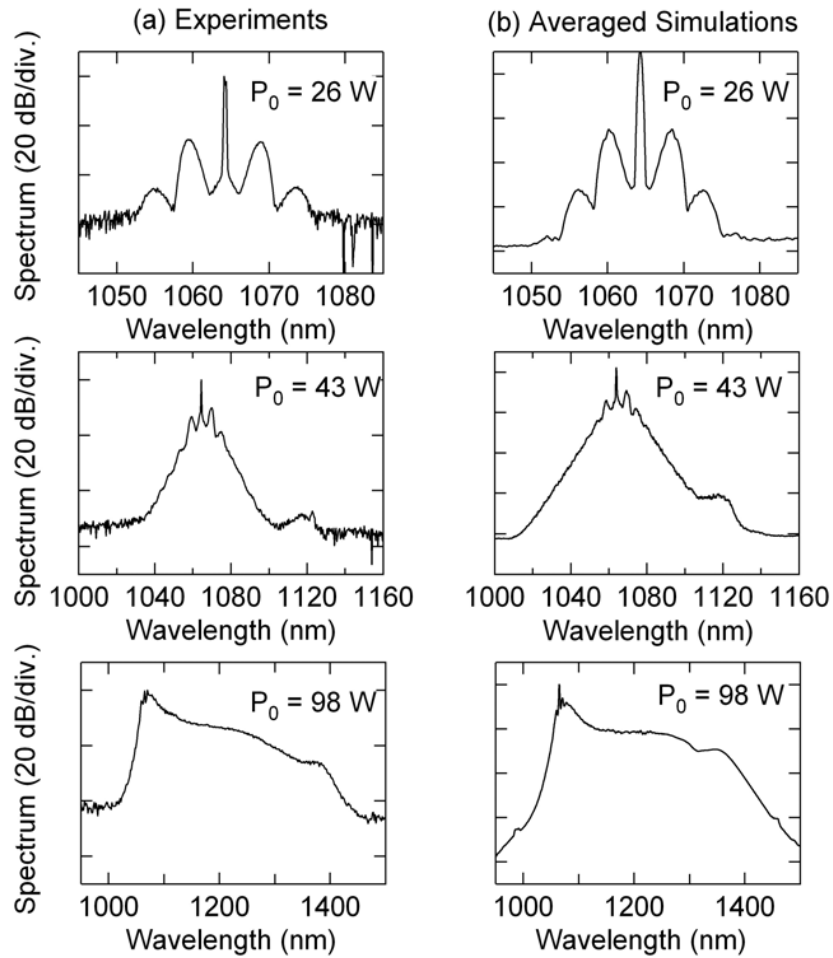

**Fig. 3** Experimental (left) and simulation (right) results for 1 ns pulses at 1064 nm injected into highly nonlinear PCF at peak powers as shown. Simulation results are averaged and convolved with a spectral resolution function matching the bandwidth of the spectrum analyzer used in the experiments (0.1 nm for 26 W results; 0.4 nm for 43 W results; 1.6 nm for 98 W results).

The agreement between experiment and simulation is very good over the full range of powers considered and the results clearly illustrate how the spectral characteristics vary significantly with power. Firstly, at the lowest power of 26 W, the spectrum consists of small number of distinct MI sidebands separated by the frequency of peak MI gain. As the power increases to 43 W, we enter an "extended MI" regime where we can still resolve sidebands close to the pump, but we also see the development of continuous low amplitude wings that appear triangular when plotted with a semi-logarithmic y-axis. The spectral width in this regime approaches 60 nm at the -40 dB level, and we note the first sign of a very low amplitude Raman peak around 1120 nm. Finally at 98 W, we see a significant increase in bandwidth to ~400 nm at the -40 dB level, and the spectrum assumes characteristics of a "fully-developed" supercontinuum [10]. We focus here on the characteristics of the extended MI spectrum at 43 W peak power. This is a regime where the increased bandwidth precludes the application of truncated sideband models, yet we still remain in an "NLSE dynamics" regime before the onset of higher-order effects. Note that at this power, the experimentally-measured sideband separation from the pump agrees very well with the 1.32 THz peak frequency of MI gain calculated for $P_0$ = 43 W and fiber parameters $\beta_2$ = -75 $ps^2 km^{-1}$ and $\gamma$ = 60 $W^{-1} km^{-1}$. The corresponding temporal period of 758 fs is much shorter than the 1 ns duration of the input pulses so that these experimental conditions are close to the ideal CW excitation case where AB theory would be expected to be valid. We also remark that we can neglect higher-order dispersion because the pump wavelength of 1064 nm is far from the fiber zero dispersion wavelength around 780 nm.

To examine how the analytic results of the AB theory above can be applied to interpret the properties of the broadened spectrum seen at 43 W, Fig. 4 presents simulation results showing in more detail how the spectrum develops with distance. The simulations are for the NLSE only with fiber parameters corresponding to experiment, and assuming a 43 W peak power CW field with a one photon per mode noise background. Simulations using the full GNLSE yield essentially identical results in this regime. The grayscale plots show results of a single numerical realization illustrating the temporal evolution of the input field over 3.9 m. The time and frequency axes of the grayscale plots are normalized relative to the frequency of peak MI gain $f_{\text{mod}} = 1.32$ THz. The spectral bandwidth first reaches its maximum extension at a distance of 3.5 m where we extract the corresponding temporal and spectral profiles.

In contrast to the narrowband modulation case considered in the Section 2 above, the use of a broadband noise seed results in significant variation in the structure of the individual temporal peaks across the extracted profile. Under these conditions, an exact comparison with the ideal AB evolution characteristics with $z$ and $T$ is not possible, but we find that the highest amplitude peaks in the temporal profile are nonetheless fitted very well by the maximally-compressed AB sub-pulses calculated from Eq. (3) using $f_{\text{mod}} = 1.32$ THz at the peak of the MI gain. This is shown explicitly in the expanded view of the shaded section in Fig. 4(a). In fact, analysis of the highest-amplitude peaks at other regions of the profile (not shown) reveals similarly good agreement with the calculated maximally-compressed AB solution, and we have confirmed that simulations using different noise seeds reveal the same characteristics. Moreover, we have also checked that the maximally-compressed AB characteristics at peak MI gain fit the highest peaks of the modulated field profiles for simulations of MI over a much wider source and fiber parameter range, indicating that this is a general characteristic of the MI process. Indeed, this can be understood in physical terms because the ABs associated with the peak MI gain frequency experience a higher growth rate and thus would be expected to make a predominant contribution to the field at the point of initial spectral expansion.

In this regard, we also remark that although the extracted profiles in Fig. 4 are taken at the first point of maximum spectral expansion, additional simulations show that the spectral and temporal characteristics do not change significantly over distances from 3.5 to 4 m, and the agreement with the AB solutions remains very good over this range. With further propagation however, the validity of a simple NLSE model breaks down as the spectral bandwidth for generated for the parameters used here approaches the peak of the Raman gain for fused silica around 13 THz.

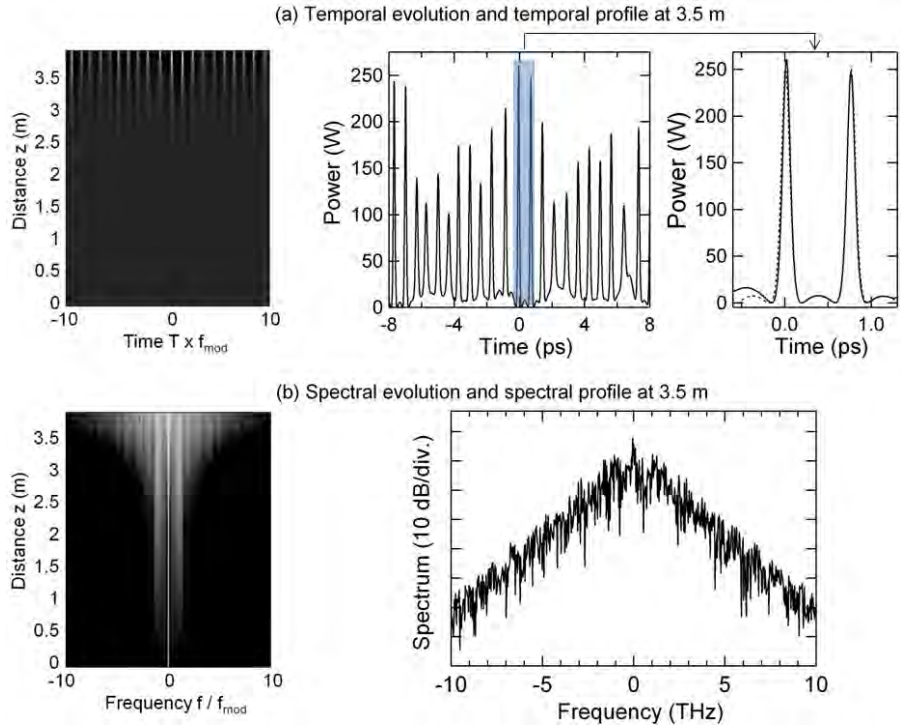

Fig 4 Single shot NLSE simulations for a CW field undergoing spontaneous MI showing: (a) temporal and (b) spectral evolution over 3.9 m. The figure also shows temporal and spectral profiles at 3.5 m. The shaded region in the temporal trace is shown in detail in the rightmost figure comparing simulations (solid line) with the AB solution calculated for a modulation frequency corresponding to peak MI gain (dashed line). The axes in the grayscale plots are normalized relative to the frequency of peak MI gain: $f_{mod}$ = 1.32 THz.

These results allow us to propose a new physical interpretation for the development of the temporal structure in spontaneous MI. Firstly, spontaneous MI leads to the development of a very large number of temporal breathers with different initial amplitudes and transverse frequencies within the bandwidth over which MI gain is observed. Secondly, the highest amplitude breathers at the point of initial spectral expansion correspond to the maximally-compressed AB solution calculated at peak MI gain which sees the highest growth rate. To our knowledge, our interpretation here is the first proposed link of spontaneous MI dynamics and the analytic properties of the AB solutions.

Significantly, we note that the highest amplitude breathers in fact possess an equivalent "soliton number" close to unity. For example, referring to Fig. 4(a) for the highest amplitude peaks with $P_{max}$ = 250 W and $\Delta\tau$ = 115 fs (FWHM), we estimate $N = [\gamma P_0 (\Delta\tau/1.763)^2 / |\beta_2|]^{1/2}$ = 0.92. This now allows us to propose a physical scenario which may be useful in interpreting the development of a fully developed broadband supercontinuum from a regime of extended MI. Specifically, we would expect that subsequent perturbations to the MI-generated breather structures would stimulate their evolution towards ideal fundamental solitons, subject of course to the fact that the chaotic initial conditions, higher-order dispersion, Raman effects and self-steepening result in different pulse accelerations such that collisions can also play a role in the subsequent dynamics. Of course, higher-order dispersion can also mediate the generation of dispersive wave radiation [10].

But the central feature of this scenario is that it suggests that, for the case of quasi-CW supercontinuum dynamics, the emergence of solitons is linked to the perturbation of high contrast pre-solitonic sub-pulses (the maximally-compressed ABs) that develop from an initial phase of breather evolution, and that are generated right across the pulse profile. This is in contrast to the process of sequential ejection of fundamental solitons from a single higher-order soliton pulse due to soliton fission, as is the case when using femtosecond pulse pumping [34, 35]. Of course, we stress that this picture needs to be complemented by more detailed studies to isolate the precise processes that stimulate the evolution of breather structures into solitons. Of particular importance will be analysis of the transition dynamics between the regime of extended MI where the field may usefully be described in terms of breathers and the fully developed supercontinuum where it has evolved into a "sea" of solitons.

In addition to providing a physical interpretation of the temporal structure of the MI field, the analytic properties of the maximally-compressed ABs also provide insight into the form of the triangular spectrum observed in Figs. 3 and 4. In fact, in the case of induced MI at a fixed modulation frequency $f_{mod}$, it is possible to obtain a general analytic description showing that the AB spectrum consists of discrete frequency sideband modes with separation $f_{mod}$, and intensities that decrease following a geometric progression [14]. For the particular case of the maximally-compressed solution of Eq. (3), the spectral mode intensities are: $S_0 = P_0 (\sqrt{2}-1)^2$ for the pump component, and $S_n = 2 P_0 (\sqrt{2}-1)^{2|n|}$ (where $n = \pm 1, \pm 2, \pm 3, \ldots$) for the sidebands. Thus the pump and sideband intensities follow the relative progression $\{I_0, I_1, I_2, I_3, I_4\ldots\} = \{1, 2, 0.3431, 0.0589, 0.0101\ldots\}$ so that there is a 3 dB increase from the pump (n=0) to the first sideband (n=±1) and then a constant decrease of $20 \log_{10}(\sqrt{2}-1) = -7.66$ dB between subsequent sidebands. This geometric progression describing the decrease in sideband amplitudes yields a characteristic triangular shape in the wings of the spectrum when plotted semi-logarithmically. These spectral characteristics are plotted in Fig. 5 for the case of the maximally-compressed AB described above with $f_{mod} = 1.32$ THz.

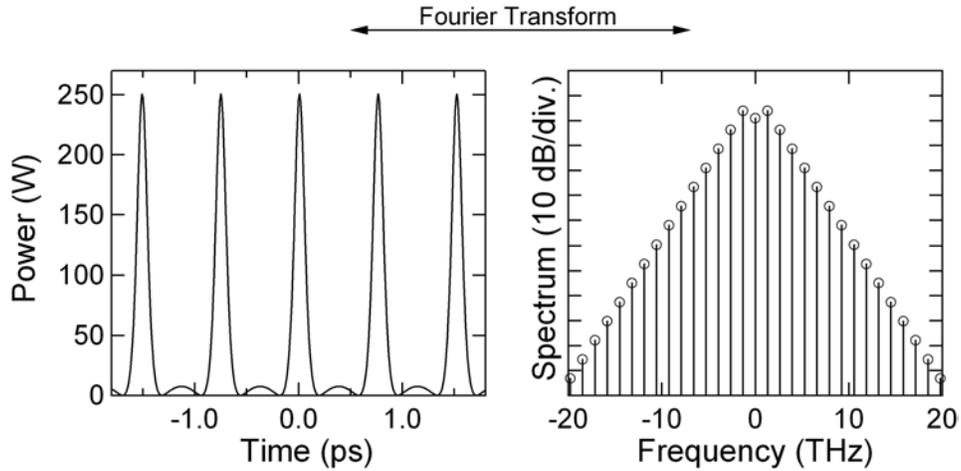

Fig 5 Time and frequency domain characteristics of the ideal maximally-compressed AB. The modulation frequency $f_{mod} = 1.32$ THz determines the spectral mode separation.

The characteristics of the AB spectrum shown in Fig. 5 suggest an immediate physical interpretation for the spectral characteristics of the spontaneous MI case seen in Fig. 4(b). However, because spontaneous MI is seeded by broadband noise, the resulting spectral structure does not contain only discrete frequency modes, but rather consists of a continuous span of frequencies (corresponding to a range of excited AB structures) without discrete gaps in the spectrum. Nonetheless, as with the time domain characteristics, because we can expect the spectral characteristics to contain a dominant component from the maximally-compressed AB solution calculated at peak MI gain, we can anticipate that the decay of the spectral wings of spontaneous MI will indeed follow the expected analytic geometric progression of the AB sideband amplitudes.

A more detailed analysis of the spontaneous MI spectrum along these lines is shown explicitly in Fig. 6. The figure compares both numerical and analytic results with the experimental results of Fig. 3 for $P_0 = 43$ W where we would expect analysis in terms of NLSE propagation and AB characteristics to be valid. The experimental results here are shown as the solid black line, and we first compare these with two sets of numerical simulations: the previous simulations shown using the full GNLSE including Raman scattering and assuming a 1 ns pulse input field (blue, short dashes), and also simulations using only the NLSE with a CW input field (red, long dashes). Note that the single-shot simulations exhibit fine-structure in the spectra similar to that shown above in Fig. 4(b), but this is not apparent in Fig. 6 which plots averages over multiple realisations (10) and also includes convolution with a 0.4 nm (0.1 THz) spectral response function to match experiment.

It is clear that the GNLSE and NLSE simulations produce very similar results, although the GNLSE result yields a slightly improved fit to experiment in showing the expected spectral asymmetry from Raman scattering. It is also clear that both sets of simulations agree quantitatively with experiment in reproducing the amplitude and slope of the wings of the MI spectrum. We can also see how (in contrast to the single shot results in Fig. 4) the averaging of the spectral structure is advantageous in more clearly showing both clear MI sidebands about the pump, as well as the clear triangular characteristic of the spectral wings.

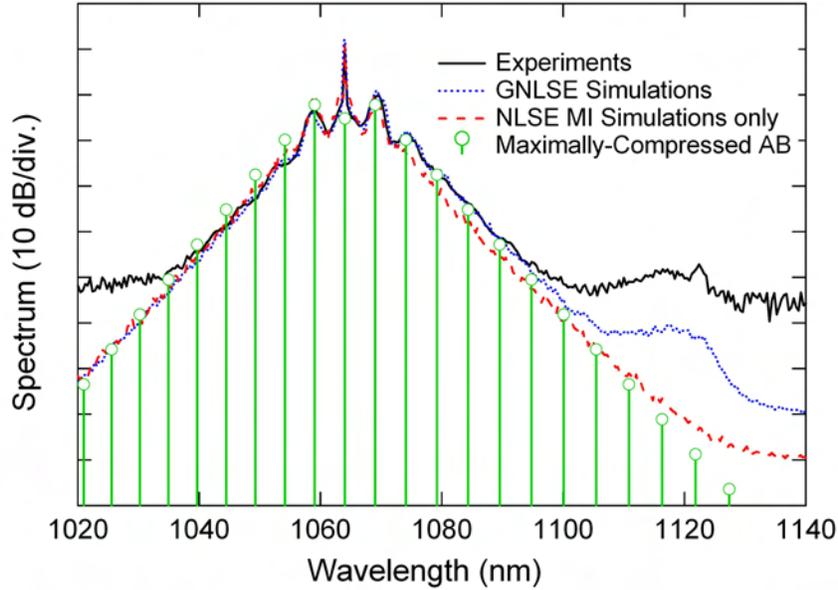

**Fig. 6** Comparison between experiments (solid black line), numerical simulations using the full GNLSE (blue dashed line), numerical simulations using the NLSE only (red dashed line), and the calculated spectrum of the maximally-compressed AB (green lines from zero).

In addition to the numerical results, the figure also shows the discrete frequency modes associated with the analytic form of the maximally-compressed AB spectrum. Because the experimental and numerical spectra are generated from broadband noise and are thus continuous, an exact quantitative comparison with the absolute amplitudes of the modes of analytic AB spectrum is not possible. This is especially the case for the relative amplitude of the pump and first sidebands where noise has a particular effect on pump depletion dynamics. Nonetheless, by normalising the amplitude of the $n = \pm 1$ sideband modes of the AB spectra to the average first sideband amplitude seen in experiments, we find that the subsequent decay of spectral intensity with frequency seen in both experiment and simulations is reproduced very well by the analytic geometric progression of the maximally-compressed AB breather.

## 4. Conclusions and Discussion

There are several major conclusions to be drawn from this work. Firstly, the analytic formalism of Akhmediev Breather propagation has been shown to accurately describe the initial stage of induced modulation instability over a wide range of conditions. Even for a non-ideal real-valued modulation amplitude, the maximally-compressed breather characteristics obtained numerically are well-described by the expected AB solution. We anticipate that these results will motivate interest into studies of the more general analytic formalism for arbitrary initial modulation conditions and equal amplitude dual mode fields [14]. Important applications of this work can also be envisaged in the optimization of experiments studying nonlinear compression and frequency comb generation [36-38].

Secondly, numerical studies of the temporal and spectral properties of spontaneous MI have shown that the characteristics of both the modulated temporal profile and the associated spectrum can be explained in terms of the development of high amplitude AB sub-pulses. Specifically, the well-known triangular form of the spontaneous MI spectrum when plotted semi-logarithmically has been explained naturally in terms of the analytic geometric progression describing the frequency dependence of the AB modal amplitudes. The analytic form of the AB spectrum has also been shown to be in very good quantitative agreement with the wings of the spontaneous MI spectrum observed in experiments studying the onset of quasi-CW supercontinuum generation.

In an optical context, an important implication of these results has been to propose a new scenario for the development of long pulse supercontinuum generation from an initial phase of spontaneous MI. Whilst the soliton fission model remains an appropriate description of supercontinuum generation for femtosecond pulse pumping where the input pulse is clearly a higher-order soliton [10], our results suggest that the onset of long pulse supercontinuum generation from spontaneous MI can be interpreted in terms of the simultaneous generation across the pulse profile of pre-solitonic Akhmediev breather structures. An important next step is to examine in more detail the dynamics of the propagation of these pre-solitonic pulses in the presence of perturbations. It is also possible that studies of breather propagation beyond the initial point of spectral expansion will reveal particular dynamical scenarios preferentially associated with optical rogue soliton trajectories that experience enhanced Raman self-frequency shift. Examining the effects of interactions and collisions between the generated breather sub-pulses may also be important in developing a more complete understanding of these instabilities and continuing the development of improved physical modelling of the supercontinuum [39]. Also of potential importance is the application of the AB formalism (or related theoretical approaches) to reconsider the way in which techniques of supercontinuum tailoring and control can be optimised more rigorously [40-41].

Finally, in a more general hydrodynamics context, our results highlight the fact that even when shot-to-shot measurements of a fluctuating field profile are not possible, the presence of high amplitude AB sub-pulses can nonetheless be inferred through quantitative analysis of the spectral wings. In fact, this technique may possibly be of general significance in removing the uncertainty into the role that nonlinear MI and breather propagation play in the development of oceanic rogue waves.

**Acknowledgements**

We thank the Institut Universitaire de France (JMD), the French Agence Nationale de la Recherche project MANUREVA ANR-08-SYSC-019 (JMD, FD, BK) and the Academy of Finland Research grants 121953 and 130099 (GG), for support. The work of NA is supported by the Australian Research Council (Discovery Project scheme DP0663216).